\documentclass[preprintnumbers, onecolumn, floatfix, preprintnumbers, amsmath, amssymb, superscriptaddress]{revtex4}

\usepackage[colorlinks,linkcolor=red,urlcolor=blue,citecolor=blue]{hyperref}
\usepackage{amssymb, amsmath, bm, dcolumn, epsf, graphicx, latexsym, mathbbol, slashed}

\begin{document}
\title{Gravitational Faraday rotation of gravitational waves by a Kerr black hole}
\author{Zhao Li}
\email{lz111301@mail.ustc.edu.cn}
\affiliation{ Department of Astronomy, University of Science and Technology of China, Hefei, Anhui 230026, China,\\ and School of Astronomy and Space Science, University of Science and Technology of China, Hefei 230026, China}
\author{Jin Qiao}
\affiliation{Purple Mountain Observatory, Chinese Academy of Sciences,
Nanjing, 210023, P.R. China,\\ and School of Astronomy and Space Science, University of Science and Technology of China, Hefei 230026, China}
\author{Wen Zhao}
\email{wzhao7@ustc.edu.cn}
\affiliation{ Department of Astronomy, University of Science and Technology of China, Hefei, Anhui 230026, China,\\ and School of Astronomy and Space Science, University of Science and Technology of China, Hefei 230026, China}
\author{Xinzhong Er}
\affiliation{South-Western Institute for Astronomy Research, Yunnan University, Kunming, Yunnan, 650000, China}

\begin{abstract}
Gravitational Faraday Rotation (GFR) is a frame-dragging effect induced by rotating massive objects, which is one of the important, yet studied characteristics of lensed gravitational waves (GWs). In this work, we calculate the GFR angle $\chi_g$ of GWs in the weak deflection limit, assuming it is lensed by a Kerr black hole (BH). We find that the GFR effect changes the initial polarization state of the lensed GW. Compared with the Einstein deflection angle, the dominant term of the rotation angle $\chi_g$ is a second-order correction to the polarization angle, which depends on the light-of-sight component of BH angular momentum. Such a rotation is tiny and degenerates with the initial polarization angle. In some critical cases, the GFR angle is close to the detection capability of the third-generation GW detector network, although the degeneracy has to be broken.
\end{abstract}

\maketitle

\section{Introduction}
According to Einstein's General Relativity (GR), a massive rotating object changes the spacetime structure and drags the inertial frame around it \cite{Chandra1983}, which causes various detectable effects on test particles \cite{weinberg,MTWbook,will}. Detecting these effects provides an excellent opportunity to study the fundamental properties and the internal structure of gravitational sources \cite{will}. For instance, Refs. \cite{Thirring1918,LenseThirring1918} described an additional precession of a gyroscope moving in a weak and slowly-rotating spacetime, precisely measured by Gravity Probe B experiment \cite{GPB2011}. The measured frame-dragging drift rate is $-37.2\pm7.2\ {\rm marcsec/yr}$, which is consistent with the prediction of GR.

Gravitational Faraday rotation (GFR) \cite{Dehnen1973,Piran1985,AnzhongWang1991} is a crucial dragging effect, a general-relativistic effect analogous to the Faraday rotation \cite{Rybicki2008}. The polarization plane of light rotates around the propagation direction when it propagates through the magnetic field. The rotation angle is proportional to the magnetic field along the line of sight (L.O.S.). In GR, the gravitational field generated by spinning massive objects is analogous to a magnetic field, interacting with electromagnetic waves (EMWs) and gravitational waves (GWs). The GFR effect of EMWs, induced by a Kerr black hole (BH), has been studied in the geometrical optical approximation \cite{Connors1977, Connors1980, Chandra1983, Ishihara1988, NouriZonoz1999, Asada2003, Sereno2004, Sereno2005, Brodutch2011, ChenBin2015, Deriglazov2021, Chakraborty2022}.

Like EMWs, the GWs are deflected by the gravitational field of massive objects, e.g., a galaxy or a BH \cite{Lawrence1971, Bray1986, Schneider1993, Sereno2006, Bozza2010, Nambu2013, Shaoqi2019}. With the discovery of GW events generated by the mergers of compact binaries \cite{GW150914, GW170817, GW190521, GWTC2019, GWTC2020}, the gravitationally-lensing effects of GW signals attract more and more attention in both observation \cite{Hannuksela2019, Abbott2021, Piorkowska2013, ShunSheng2018} and theory \cite{Bray1986,Baraldo1999,Takahashi2003,Shaoqi2019}. In addition to the lensing effects such as multiple images, magnification, phase shift, interference, and diffraction \cite{Nakamura1997, LiangDai2018, Meena2019, Mishra2021, Cheung2021, Nakamura1999, Takahashi2003,GuoXiao2020, Baraldo1999, XikaiShan2022}, the polarization plane of GW can be dragged by the angular momentum of the lens, i.e., the GFR of GWs \cite{AnzhongWang1991, XilongFan2017, kailiao2017, Shaoqi2019}. The lensing effects of GWs have been explicitly studied in the geometric optics or diffraction theory framework \cite{Born1959}, except for the GFR effect. In this work, we will present the explicit result of the GFR angle of GWs induced by a Kerr BH by solving the geodesic equations of massless particles \cite{Ishihara1988, Chandra1983, Bray1986, Sereno2006}. We focus on the GFR effect in the weak deflection limit (WDL) of strong gravitational lensing. The geometric optical approximation is valid since the wavelength of the GW is assumed to be not comparable with the gravitational radius of the lens in our studies.

The organization of this article is as follows. Section \ref{sec-2} is a brief introduction to the Walker-Penrose theorem for Kerr spacetime. Then, we describe a suitable definition of the GW metric in Section \ref{sec-3} and present our derivation of the GFR angle of GW in Section \ref{sec-4}. Finally, Section \ref{sec-5} shows the evaluation result of the GFR angle of GW. We introduce the observational effect of GFR in gravitationally-lensed GW signals in Section \ref{sec-6}. Section \ref{sec-7} concludes and discusses the main results of this paper. In this article, we will use units $c=G=1$, where $c$ is the speed of light in the vacuum, and $G$ is the gravitational constant.

\section{Walker-Penrose Theorem}
\label{sec-2}
A Kerr BH is fully described by its mass $M$ and angular momentum $a$. In Boyer-Lindquist coordinates $(t,r,\theta,\varphi)$ \cite{BoyerLindquist1967}, the metric of Kerr spacetime $g_{\mu\nu}$ \cite{Kerr1963} is 
\begin{equation}
\label{kerr}
ds^2=-dt^2+\frac{\rho^2}{\Delta}dr^2+\rho^2d\theta^2
+(r^2+a^2)\sin^2\theta d\varphi^2+\frac{2Mr}{\rho^2}\left(a\sin^2\theta d\varphi-dt\right)^2,
\end{equation}
where $\Delta=r^2-2Mr+a^2$ and $\rho^2=r^2+a^2\cos^2\theta$. Let $x^{\mu}=x^{\mu}(\tau)$ be an affine-parameterized null geodesic and $k^{\mu}$ be the tangent vector of $x^{\mu}(\tau)$, $k^{\mu}\equiv\partial x^{\mu}/\partial\tau$, where $\tau$ is proper time. $k^{\mu}$ is light-like and parallel transported along the geodesic, i.e. $|\bm{k}|^2=g_{\mu\nu}k^{\mu}k^{\nu}=0$ and $k^{\alpha}\nabla_{\alpha}k_{\mu}=0$. The vector $f^{\mu}$ is assumed to be orthogonal to $k^{\mu}$, and parallel transports along the null geodesic, i.e., $\bm{k}\cdot\bm{f}=g_{\mu\nu}k^{\mu}f^{\nu}=0$ and $k^{\alpha}\nabla_{\alpha}f_{\mu}=0$. Since the Kerr geometry is of Petrov type D \cite{Chandra1983}, one can define Walker-Penrose (WP) conserved quantity,
\begin{equation}
\label{WP}
K_{\rm WP}=(A-iB)(r-ia\cos\theta)=K_{1}+iK_{2},
\end{equation}
WP conserved quantity (\ref{WP}) satisfies the WP theorem \cite{WalkerPenrose1970}, 
\begin{equation}
\label{WP-theorem}
k^{\mu}\nabla_{\mu}K_{\rm WP}=0,
\end{equation}
where $A\equiv(\bm{k}\cdot\bm{l})(\bm{f}\cdot\bm{n})-(\bm{k}\cdot\bm{n})(\bm{f}\cdot\bm{l})$ and $iB\equiv(\bm{k}\cdot\bm{m})(\bm{f}\cdot\bar{\bm{m}})-(\bm{k}\cdot\bar{\bm{m}})(\bm{f}\cdot\bm{m})$ \cite{Chandra1983}. $\{\bm{l},\bm{n},\bm{m},\bar{\bm{m}}\}$ is the Newman-Penrose tetrad of Kerr BHs \cite{Kinnersley1969}. $K_{1}$ and $K_2$ are the real and imaginary parts of $K_{\rm WP}$, respectively. WP theorem has been used to predict the GFR of EMWs by treating $f^{\mu}$ as the EMW polarization vector \cite{Ishihara1988, Brodutch2011, NouriZonoz1999}. One can always set $f^{t}=0$ without loss of generality. The radiation gauge causes $\bm{f}$ to always be in the polarization plane. WP theorem constrains the evolution of polarization vector along a null geodesic. Thus, the polarization vector $\bm{f}$ rotates on the polarization plane during propagation. 

Far from the BH, the WP conserved quantity (\ref{WP}) is 
\begin{equation}
\label{K1K2}
\left\{
\begin{aligned}
K_{1}&=r\left[-\beta f^{\theta}-(\gamma\sin\theta)f^{\varphi}\right]\frac{k^{r}}{|k^{r}|}\\
K_{2}&=r\left[+\gamma f^{\theta}-(\beta\sin\theta)f^{\varphi}\right]
\end{aligned}
\right.,
\end{equation}
where $\beta\equiv\sqrt{\eta+a^2\cos^2\theta-\xi^2\cot^2\theta}{k^{\theta}}/{|k^{\theta}|}$ and $\gamma\equiv\xi\csc\theta-a\sin\theta$. In (\ref{K1K2}), $\eta$ is the $z$-component of angular momentum, and $\xi$ is the Carter constant of a massless particle \cite{Carter1968}. Since $K_{1}$ and $K_{2}$ are finite, $f^{\theta}$ and $f^{\varphi}$ approach $0$ when $r$ is large, and $f^{r}$ also vanishes due to the gauge condition $\bm{k}\cdot\bm{f}=0$. Therefore, it is convenient to introduce a unit transverse vector $\bm{\mathcal{E}}$ with components,
\begin{equation}
\label{E-def}
\mathcal{E}^{r}\equiv f^{r},\quad
\mathcal{E}^{\theta}\equiv-rf^{\theta},\quad \mathcal{E}^{\varphi}\equiv-r\sin\theta f^{\varphi}.
\end{equation}
$\mathcal{E}^{r}$ vanishes for large $r$. Thus, $\bm{\mathcal{E}}$ is a vector tangent to a large sphere, with two orthogonal components $\mathcal{E}^{\theta}$ and $\mathcal{E}^{\varphi}$, and conserved norm $[(\mathcal{E}^{\theta})^2+(\mathcal{E}^{\varphi})^2]^{1/2}=(K_{1}^2+K_{2}^2)^{1/2}$. This equality means that the energy carried by EMWs is conserved during propagation in geometrical optics approximation. Therefore, the asymptotic form of WP conserved quantity is expressed in $\mathcal{E}^{\theta}$ and $\mathcal{E}^{\varphi}$,
\begin{equation}
\label{K-asy}
\left\{
\begin{aligned}
\frac{k^{r}}{|k^{r}|}K_{1}
&=+\beta\mathcal{E}^{\theta}+\gamma\mathcal{E}^{\varphi}\\
K_{2}&=-\gamma\mathcal{E}^{\theta}+\beta\mathcal{E}^{\varphi}
\end{aligned}\right..
\end{equation}
The WP theorem requires that $K_{1}$ and $K_{2}$ have the same values along the graviton geodesic. The conservation of $K_{\rm WP}$ provides an exact translation relation of $\bm{\mathcal{E}}$ between the source and detector positions. This transformation is given by Ref. \cite{Ishihara1988} and the following Eq.(\ref{E-trans}) in Section \ref{sec-4}.

\section{Definition of GW Metric}
\label{sec-3}
The WP theorem constrains the evolution of the unit transverse vector $\bm{\mathcal{E}}$ and can also constrain the evolution of GWs. In this section, the GW metric tensor $h^{ij}$ is defined to prepare to derive the GFR angle of gravitationally-lensed GWs by a Kerr BH in the next section.

\begin{figure}
\centering
\includegraphics[width=8cm]{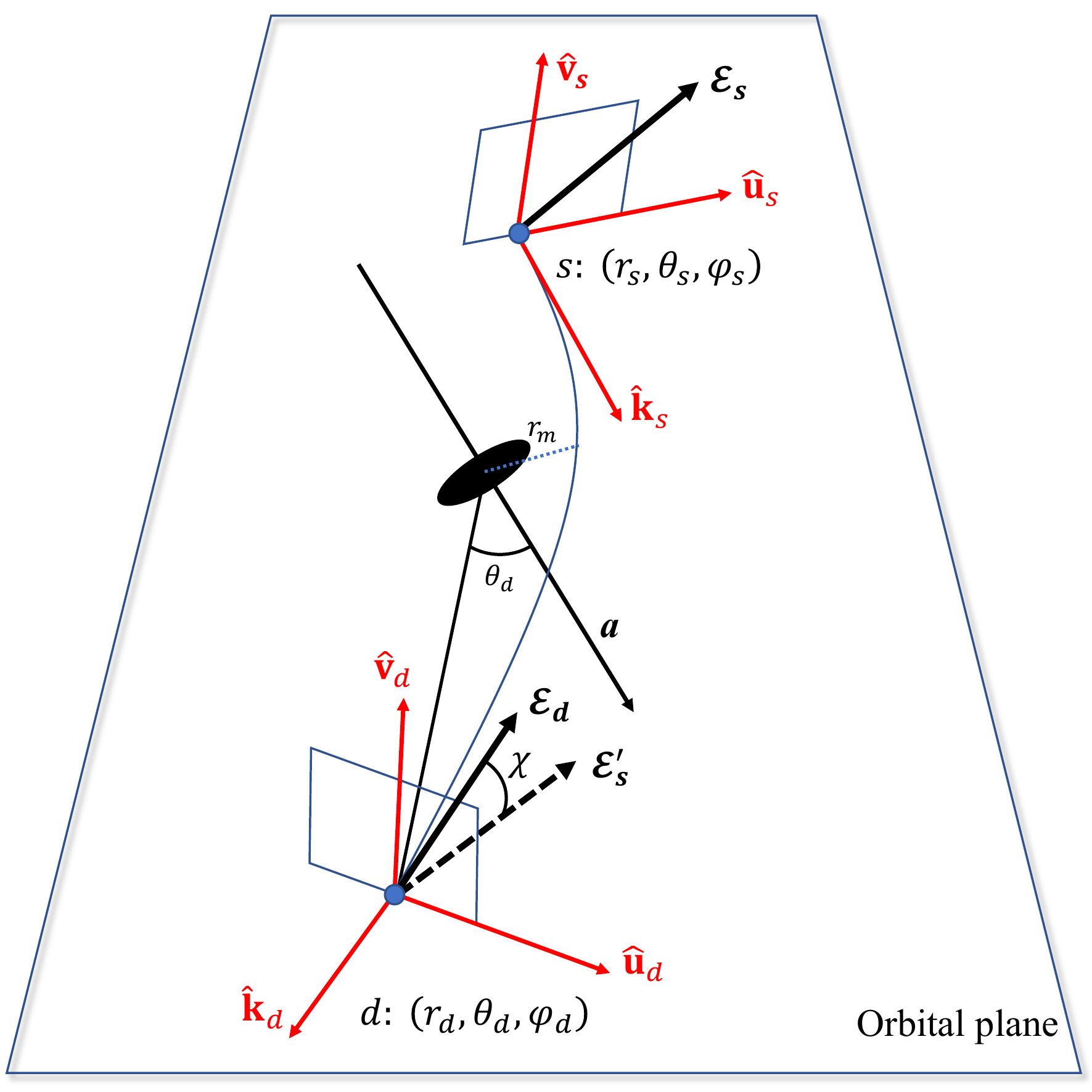}
\caption{Demonstration of GFR of lensed GW system. The gravitons emitted from a source at $(r_{s},\theta_{s},\varphi_{s})$ pass near a Kerr BH with angular momentum $\bm{a}$, and then arrive at a detector at $(r_{d},\theta_{d},\varphi_{d})$.}
\label{fig:geometry figure}
\end{figure}

Firstly, the metric tensor $F^{\mu\nu}$ is defined based on vector $\bm{f}$ as follows,
\begin{equation}
\label{F-def}
F^{\mu\nu}\equiv f^{\mu}f^{\nu},
\ (|\bm{f}|\ll1).
\end{equation}
Obviously, $F^{\mu\nu}$ is symmetric, orthogonal to the wave vector $\bm{k}$, and parallel transported along null geodesics. Thus, we conclude that the definition (\ref{F-def}) satisfies the requirements of a metric tensor for the GW. Secondly, we introduce the traceless condition using transverse-traceless (TT) projector \cite{Maggiore2008} $\Lambda_{ij,kl}(\hat{\mathbf{k}})=P_{ik}P_{jl}-(1/2)P_{ij}P_{kl}$ with $P_{ij}(\hat{\mathbf{k}})=\eta_{ij}-\hat{k}_{i}\hat{k}_{j}$, where $\eta_{ij}=\text{diag}(1,r^2,r^2\sin^2\theta)$ is the spatial part of a flat background metric at an infinite distance, $\hat{\mathbf{k}}\equiv\mathbf{k}/|\mathbf{k}|$, and $\mathbf{k}$ is the spatial part of the wave vector $\bm{k}$. Considering a Schwarzschild lens, the particle orbit will be in the BH's equatorial plane, and so will the background source and the detector. This plane is also called the source-lens-detector plane. This conclusion is generally invalid for a Kerr lens. However, under the WDL approximation, the trajectories of gravitons deviate sufficiently small from the source-lens-detector plane, which is not generally the equatorial plane of Kerr BH \cite{Bray1986,Ishihara1988,Brodutch2011}. Therefore, the source-lens-detector plane can safely regarded as the orbital plane, presented in Fig.\,\ref{fig:geometry figure}. The source and detector positions are described as coordinates $(r_{s},\theta_{s},\varphi_{s})$ and $(r_{d},\theta_{d},\varphi_{d})$, which are sufficiently distant from the BH. Therefore, these two wave vectors are approximated as $\hat{\mathbf{k}}_{s}=(-1,0,0)$ and $\hat{\mathbf{k}}_{d}=(1,0,0)$ at these two positions. The subscripts $s$ and $d$ represent the corresponding quantities evaluated at the source and detector positions, respectively. TT projector gives an equivalent metric tensor of (\ref{F-def}), $H^{ij}$, in TT gauge,
\begin{equation}
\label{H-def}
H^{ij}
=\Lambda_{ij,kl}(\hat{\mathbf{k}}_{s,d})F^{kl}
=\left(\begin{array}{cc}
\frac{1}{2}\left(F^{\theta\theta}
-F^{\varphi\varphi}\sin^2\theta\right)
& F^{\theta\varphi}\\
F^{\theta\varphi}
& \frac{1}{2}\left(F^{\varphi\varphi}
-F^{\theta\theta}\csc^2\theta\right)\\
\end{array}\right),
\end{equation}
which is tangent to an infinitely large sphere. The $r$-component is removed for simplification. One can find that tensor $H^{ij}$ (\ref{H-def}) only has two independent components, $H^{\theta\theta}$ and $H^{\theta\varphi}$. As a result, the tensor $H^{ij}$ describes a beam of GW propagating along with vector $\hat{\mathbf{k}}_{s}$ or $\hat{\mathbf{k}}_{d}$ at infinity. Furthermore, after transforming $H^{ij}$ into a Cartesian system, the transformed tensor $\tilde{H}^{ij}$ (where $i,j=x,y,z$, $x\equiv r\sin\theta\cos\varphi, y\equiv r\sin\theta\sin\varphi, z\equiv r\cos\theta$) describes GWs propagating along with the direction $(\sin\theta\cos\varphi,\sin\theta\sin\varphi,\cos\theta)$ in TT gauge.

In (\ref{E-def}), the normalized vector $\bm{\mathcal{E}}$ is introduced. A similar operation is applied to the metric tensors, which is convenient for constructing basis vectors and projecting metric tensors. Generally, a vector $\bm{f}$ can be written as $\bm{f}=f^{i}\bm{e}_{i}=\mathcal{E}^{i}\hat{\bm{e}}_{i}$, where $\bm{e}_{i}=\bm{e}_{r},\ \bm{e}_{\theta},\ \bm{e}_{\varphi}$ are basis vectors and $\hat{\bm{e}}_{r}\equiv\bm{e}_{r},\ \hat{\bm{e}}_{\theta}\equiv-(1/r)\bm{e}_{\theta},\ \hat{\bm{e}}_{\varphi}\equiv-(1/r\sin\theta)\bm{e}_{\varphi}$ are unit basis vectors in the spherical coordinate system. Similarly, a tensor $\bm{H}$ can generally be written as  $\bm{H}=H^{ij}\bm{e}_{i}\bm{e}_{j}=h^{ij}\hat{\bm{e}}_{i}\hat{\bm{e}}_{j}$. Combining the above discussion, we define a new metric tensor with component, $h^{\theta\theta}\equiv r^2H^{\theta\theta}$, $h^{\theta\varphi}\equiv r^2\sin\theta H^{\theta\varphi}$, and $h^{\varphi\varphi}\equiv r^2\sin^2\theta H^{\varphi\varphi}$, which is expressed as
\begin{equation}
\label{h-def}
h^{ij}=\left(\begin{array}{cc}
h^{\theta\theta} & h^{\theta\varphi}\\
h^{\theta\varphi} & h^{\varphi\varphi}
\end{array}\right)
=\left(\begin{array}{cc}
\frac{1}{2}\left[(\mathcal{E}^{\theta})^2-(\mathcal{E}^{\varphi})^2\right] &
\mathcal{E}^{\theta}\mathcal{E}^{\varphi}\\
\mathcal{E}^{\theta}\mathcal{E}^{\varphi} &
-\frac{1}{2}\left[(\mathcal{E}^{\theta})^2-(\mathcal{E}^{\varphi})^2\right]
\end{array}\right).
\end{equation}
Note that, the metric $h^{ij}$ (\ref{h-def}) is an equivalent expression of $F^{ij}$ (\ref{F-def}) in TT gauge and is expressed in unit basis vectors.

In conclusion, the definition of the GW metric  comes from the direct product of two transverse vectors and satisfies all the physical properties of GW. TT gauge is then applied to remove the residual degree of freedom. The Cartesian-coordinate form of metric tensor is given in (\ref{h-def}).

\section{Gravitational Faraday Rotation}
\label{sec-4}
In this section, we derive the GFR angle of the lensed GWs and reproduce the previous results of the EMW case in Ref. \cite{Ishihara1988} for comparison. To find the transformation matrix for polarization $(h^{\theta\theta},h^{\theta\varphi})$ between the source and the detector positions, we firstly calculate the transformation matrix of the unit transverse vector $(\mathcal{E}^{\theta},\mathcal{E}^{\varphi})$. The WP theorem conserves $K_{1}$ and $K_{2}$ at the source and detector positions, calculated from (\ref{K-asy}). The conservation gives $\beta_{s}\mathcal{E}_{s}^{\theta}+\gamma_{s}\mathcal{E}_{s}^{\varphi}=-(\beta_{d}\mathcal{E}_{d}^{\theta}+\gamma_{d}\mathcal{E}_{d}^{\varphi})$ and $-\gamma_{s}\mathcal{E}_{s}^{\theta}+\beta_{s}\mathcal{E}_{s}^{\varphi}=-\gamma_{d}\mathcal{E}_{d}^{\theta}+\beta_{d}\mathcal{E}_{d}^{\varphi}$. The electric vectors $\bm{\mathcal{E}}$ at the source and detector positions are related through a transformation matrix \cite{Ishihara1988}
\begin{equation}
\label{E-trans}
\left(
\begin{array}{c}
\mathcal{E}^{\theta}_{d}\\
\mathcal{E}^{\varphi}_{d}
\end{array}
\right)
=\bm{\mathcal{T}}
\left(
\begin{array}{c}
\mathcal{E}^{\theta}_{s}\\
\mathcal{E}^{\varphi}_{s}
\end{array}
\right),
\end{equation}
which can be written explicitly as
\begin{equation}
\bm{\mathcal{T}}
=-\frac{1}{\beta_{d}^2+\gamma_{d}^2}
\left(\begin{array}{cc}
\beta_{d}\beta_{s}-\gamma_{d}\gamma_{s} 
& \beta_{d}\gamma_{s}+\gamma_{d}\beta_{s}\\
\beta_{d}\gamma_{s}+\gamma_{d}\beta_{s} 
& \gamma_{d}\gamma_{s}-\beta_{d}\beta_{s}
\end{array}\right).
\end{equation}
Furthermore, the transformation matrix $\bm{\mathcal{T}}$ can be simply rewritten as
\begin{equation}
\label{R-def}
\bm{\mathcal{T}}=\frac{1}{\sqrt{1+x^2}}
\left(\begin{array}{cc}1 & -x\\-x & -1\end{array}\right).   
\end{equation}
through a new symbol $x$, defined as
\begin{equation}
\label{x-def}
x\equiv \frac{\gamma_{s}\beta_{d}+\beta_{s}\gamma_{d}}{\gamma_{s}\gamma_{d}-\beta_{s}\beta_{d}}.
\end{equation}
The quantity $x$ (\ref{x-def}) depends on the motion constants and angular coordinates of the source. In this new form (\ref{R-def}), the equality $\beta^2+\gamma^2=\eta+(\xi-a)^2$ has been considered. The matrix $\bm{\mathcal{T}}$ describes the full change in the transverse vector. Only two independent components remain when applying radiation gauge for transverse vector $\bm{\mathcal{E}}$. The change between the final state $(\mathcal{E}_{d}^{\theta},\mathcal{E}_{d}^{\varphi})$ and the initial state $(\mathcal{E}_{s}^{\theta},\mathcal{E}_{s}^{\varphi})$ represents a rotation on the polarization plane, determined by WP theorem.

\subsection{GW case}
Above transformation (\ref{E-trans}) and the definition of $h^{ij}$ (\ref{h-def}) give the transformation of the independent component of GW metric between the source and detector positions,
\begin{equation}
\label{h-trans}
\left(\begin{array}{c} h^{\theta\theta}_{d} \\ h^{\theta\varphi}_{d}\end{array}\right)
=\bm{\mathcal{S}}
\left(\begin{array}{c} h^{\theta\theta}_{s}\\
h^{\theta\varphi}_{s}\end{array}\right),
\end{equation}
with transformation matrix
\begin{equation}
\label{S-matrix}
\bm{\mathcal{S}}=\frac{1}{1+x^2}
\left(\begin{array}{cc} 1-x^2 & -2x\\ -2x & x^2-1\end{array}\right).
\end{equation}
The matrix $\bm{\mathcal{S}}$ describes the total change in the $\theta\theta$ and $\theta\varphi$ components of the metric tensor.

To present the measurement of the GW polarization, we construct a set of basis vectors as a reference frame, $\hat{\mathbf{u}}$ and $\hat{\mathbf{v}}$, and project metric tensor $h^{ij}$ into them. The vectors $\hat{\mathbf{u}}$ and $\hat{\mathbf{v}}$ need to be orthogonal to $\hat{\mathbf{k}}$ or, equivalently, on the polarization plane. Without loss of generality, $\hat{\mathbf{v}}$ is assumed to be orthogonal to the orbital plane and $\hat{\mathbf{u}}=\hat{\mathbf{v}}\times\hat{\mathbf{k}}$ (Fig.\,\ref{fig:geometry figure}). As mentioned in section \ref{sec-3}, the wavevectors $\hat{\mathbf{k}}_s$ and $\hat{\mathbf{k}}_d$ are on the orbital plane. The normal vector of the orbital plane, $\mathbf{v}$, is then determined approximately by $\mathbf{v}=\hat{\mathbf{k}}_{d}\times\hat{\mathbf{k}}_{s}$ and the basis vector is $\hat{\mathbf{v}}=(v^{r},r v^{\theta},r\sin\theta v^{\varphi})$. Another basis vector $\hat{\mathbf{u}}$ is given by $\hat{\mathbf{u}}_{s}=(0,-\hat{v}_{s}^{\varphi},\hat{v}_{s}^{\theta})$ at the source position, and $\hat{\mathbf{u}}_{d}=(0,\hat{v}_{d}^{\varphi},-\hat{v}_{d}^{\theta})$ at the detector position. For simplification, we define $n_{s}\equiv\hat{v}_{s}^{\varphi}/\hat{v}_{s}^{\theta}$, $n_{d}\equiv\hat{v}_{d}^{\varphi}/\hat{v}_{d}^{\theta}$ and write them as following form,
\begin{equation}
\label{nsd}
n_{s}=\sin\theta_{s}
\frac{+\cot\theta_{d}-\cot\theta_{s}\cos(\varphi_{d}-\varphi_{s})}
{\sin(\varphi_{d}-\varphi_{s})}\quad\text{and}\quad
n_{d}=\sin\theta_{d}
\frac{-\cot\theta_{s}+\cot\theta_{d}\cos(\varphi_{d}-\varphi_{s})}
{\sin(\varphi_{d}-\varphi_{s})},
\end{equation}
to represent $\hat{\mathbf{v}}_{s}$ and $\hat{\mathbf{v}}_{d}$, equivalently. The polarization tensors are then constructed as $e_{+}^{ij}\equiv\hat{u}^{i}\hat{u}^{j}-\hat{v}^{i}\hat{v}^{j}$ and $e_{\times}^{ij}\equiv\hat{u}^{i}\hat{v}^{j}+\hat{v}^{i}\hat{u}^{j}$ \cite{Maggiore2008},
which are expressed explicitly as
\begin{equation}
\label{epec-expr}
\begin{aligned}
e_{+}^{ij}&=\left(\begin{array}{cc}
(\hat{u}^{\theta})^2-(\hat{v}^{\theta})^2 & \hat{u}^{\theta}\hat{u}^{\varphi}-\hat{v}^{\theta}\hat{v}^{\varphi}\\
\hat{u}^{\theta}\hat{u}^{\varphi}-\hat{v}^{\theta}\hat{v}^{\varphi} & (\hat{u}^{\varphi})^2-(\hat{v}^{\varphi})^2 \\
\end{array}\right),\quad\text{and}\quad
e_{\times}^{ij}&=\left(\begin{array}{cc}
2\hat{u}^{\theta}\hat{v}^{\theta} & \hat{u}^{\theta}\hat{v}^{\varphi}+\hat{v}^{\theta}\hat{u}^{\varphi}\\
\hat{u}^{\theta}\hat{v}^{\varphi}+\hat{v}^{\theta}\hat{u}^{\varphi} & 2\hat{u}^{\varphi}\hat{v}^{\varphi}
\end{array}\right),
\end{aligned}
\end{equation}
respectively. The polarization tensors are normalized and orthogonal to each other, $e_{P}^{ij}e_{P'}^{ij}=2\delta_{PP'}$, where subscripts $_{P,P'}=+,\times$. The metric $h^{ij}$ is decomposed into plus and cross modes, $h^{ij}=h^{+}e_{+}^{ij}+h^{\times}e_{\times}^{ij}$. The polarization modes are $h^{+}=(1/2)h^{ij}e_{ij}^{+}$ and $h^{\times}=(1/2)h^{ij}e_{ij}^{\times}$, respectively. Rearranging it as a matrix form, we obtain
\begin{equation}
\label{h-projection}
\left(\begin{array}{c}h^{+}\\ h^{\times}\end{array}\right)
=\bm{\mathcal{K}}
\left(\begin{array}{c}h^{\theta\theta} \\ h^{\theta\varphi} \end{array}\right)
\end{equation}
with projection matrix
\begin{equation}
\label{K-matrix}
\bm{\mathcal{K}}=\left(\begin{array}{cc}
(\hat{u}^{\theta})^2-(\hat{v}^{\theta})^2 & \hat{u}^{\theta}\hat{u}^{\varphi}-\hat{v}^{\theta}\hat{v}^{\varphi}\\
\hat{u}^{\theta}\hat{v}^{\theta}-\hat{u}^{\varphi}\hat{v}^{\varphi} & \hat{u}^{\varphi}\hat{v}^{\theta}+\hat{u}^{\theta}\hat{v}^{\varphi}
\end{array}\right).
\end{equation}
The metric tensor $h^{ij}$ projects onto two basis tensors through the $\bm{\mathcal{K}}$ matrix, and two polarization modes $h^{+}$ and $h^{\times}$ of GWs are obtained.

Combining transformation (\ref{h-trans}) and projection (\ref{h-projection}), one can obtain the rotation relation of the polarization modes from the source to the detector,
\begin{equation}
\label{hphc-trans}
\begin{aligned}
\left(\begin{array}{c}h^{+}_{d}\\ h^{\times}_{d} \end{array}\right)
=\bm{\mathcal{K}}_{d}
\left(\begin{array}{c}h^{\theta\theta}_{d}\\ h^{\theta\varphi}_{d}\end{array}\right)
=\bm{\mathcal{K}}_{d}\bm{\mathcal{S}}
\left(\begin{array}{c}h^{\theta\theta}_{s}\\ h^{\theta\varphi}_{s}\end{array}\right)
=\bm{\mathcal{K}}_{d}\bm{\mathcal{S}}\bm{\mathcal{K}}_{s}^{-1}
\left(\begin{array}{c}h^{+}_{s}\\ h^{\times}_{s}\end{array}\right)
\end{aligned}.
\end{equation}
The rotation of a polarization plane (denoted by an angle $\chi_{g}$) transforms the polarization modes of GWs by $h^{+}\rightarrow h^{+}\cos2\chi_{g}-h^{\times}\sin2\chi_{g}$ and $h^{\times}\rightarrow h^{+}\sin2\chi_{g}+h^{\times}\cos2\chi_{g}$ \cite{Maggiore2008}, where the factor $2$ comes from the spin-$2$ symmetry of the graviton. Thus, the production of three matrices in (\ref{hphc-trans}) reduces to a rotation matrix,
\begin{equation}
\label{h-rotation}
\bm{\mathcal{R}}\equiv\bm{\mathcal{K}}_{d}\bm{\mathcal{S}}\bm{\mathcal{K}}_{s}^{-1}
=\left(\begin{array}{cc}\cos 2\chi_{g} & -\sin 2\chi_{g}\\ \sin 2\chi_{g} & \cos 2\chi_{g}\end{array}\right),
\end{equation}
and the GFR angle $\chi_{g}$ of GWs is given by
\begin{equation}
\label{chi-g}
\chi_{g}=\frac{1}{1+x^2}\left(\frac{X-x}{1+X^2}-\frac{X}{1+X^2}x^2+Yx\right),
\end{equation}
where 
\begin{equation}
\label{XY-def}
X\equiv-\frac{n_{s}+n_{d}}{1-n_{s}n_{d}}
\quad
{\rm and} \quad
Y\equiv\frac{(n_{s}+n_{d})^2}{(1+n_{s}^2)(1+n_{d}^2)}.
\end{equation}
The quantities $X$ and $Y$ depend on the choice of basis vectors $\hat{\mathbf{u}}$ and $\hat{\mathbf{v}}$. The matrix $\bm{\mathcal{R}}$ is real orthogonal, ensuring the equality $(h_{s}^{+})^2+(h_{s}^{\times})^2=(h_{d}^{+})^2+(h_{d}^{\times})^2$, i.e. the energy carried by GW is conserved during propagation. 

We should mention that the WP theorem (\ref{WP-theorem}) is independent of the reference frame, governing the GFR effect of light and GW rays. Therefore, the polarization angle $\psi$ is also frame-dependent, describing the polarization states of photons and gravitons. Thus, defining the two standard polarization directions is necessary to investigate the initial and final polarization states. Therefore, there are frame-dependent terms ($n_{s}$ and $n_{d}$) in the final result (\ref{chi-g}). But we only focus on the GFR angle, the difference between the initial and final polarization state, which is frame-independent.

In the WDL assumption, $\hat{\mathbf{k}}_{s}$ and $\hat{\mathbf{k}}_{d}$ are approximately coplanar. For convenience, we follow the method proposed by Ref. \cite{Ishihara1988}, requiring the $\hat{\mathbf{v}}$-axis to be perpendicular to the orbital plane, or equivalently, the $\hat{\mathbf{u}}$-axis to be in the orbital plane, as shown in Fig.\,\ref{fig:geometry figure}. The GFR effect disappears for a nonspinning lens. One will find that the angle between the polarization direction and the $\hat{\mathbf{u}}$-axis is constant during propagation. Thus, the initial and final polarization angles are equal, and the GFR angle is zero. For the Kerr lens, the GFR angle is obtained in the reference frame shown in Fig.\,\ref{fig:geometry figure}. In another frame (for example, one requires a nonzero angle between the $\hat{\mathbf{u}}$-axis and orbital plane), the measured polarization angle is different. However, we can verify that the GFR angle remains unchanged.

In order to prove the frame covariance of GFR angle, we rotate the $(\hat{\mathbf{u}}-\hat{\mathbf{v}})$ plane around $\hat{\mathbf{k}}_s$ and $\hat{\mathbf{k}}_d$ by an angle $\phi$. At this time, the symbols $n_s$ and $n_d$ will be transformed as follows,
\begin{equation}
\begin{aligned}
n_{s}\rightarrow n_{s}(\phi)=&\frac{\cos\phi
[\cos\theta_{d}\sin\theta_{s}
-\sin\theta_{d}\cos\theta_{s}\cos(\varphi_{d}-\varphi_{s})]
+\sin\phi\sin\theta_{d}\sin (\varphi_{d}-\varphi_{s})}
{-\sin\phi[\sin\theta_{s}\cos\theta_{d}
-\sin\theta_{d}\cos\theta_{s}
\cos (\varphi_{d}-\varphi_{s})]
+\cos\phi\sin\theta_{d}\sin(\varphi_{d}-\varphi_{s})},\\
n_{d}\rightarrow n_{d}(\phi)=&\frac{\cos\phi
[\sin\theta_{s}\cos\theta_{s}
-\cos\theta_{d}\sin\theta_{s}
\cos(\varphi_{d}-\varphi_{s})]
+\sin\phi\sin\theta_{s}
\sin(\varphi_{d}-\varphi_{s})}
{+\sin\phi[\sin\theta_{s}\cos\theta_{s}
-\sin\theta_{s}\cos\theta_{s}
\cos (\varphi_{d}-\varphi_{s})]
-\cos\phi\sin\theta_{s}\sin (\varphi_{d}-\varphi_{s})}.
\end{aligned}
\end{equation}
According to (\ref{XY-def}), under the rotation transformation, $X$ and $Y$ will become
\begin{equation}
\label{Xphi}
X\rightarrow X(\phi)=-\frac{(\cos\theta_{d}-\cos\theta_{s})
\sin(\varphi_{d}-\varphi_{s})}
{\cos\varphi_{d}\cos\varphi_{s}
(\cos\theta_{d}\cos\theta_{s}-1)
+\sin\varphi_{d}\sin\varphi_{s}
(\cos\theta_{d}\cos\theta_{s}-1)
+\sin\theta_{d}\sin\theta_{s}},
\end{equation}
and
\begin{equation}
\label{Yphi}
Y\rightarrow Y(\phi)=\left[\frac{(\cos\theta_{d}-\cos\theta_{s})
\sin(\varphi_{d}-\varphi_{s})}
{\sin\theta_{d}\sin\theta_{s}\sin\varphi_{d}\sin\varphi_{s} +\sin\theta_{d}\sin\theta_{s}\cos\varphi_{d}\cos\varphi_{s}
+\cos\theta_{d}\cos\theta_{s}-1}\right]^2.
\end{equation}
Eqs.(\ref{Xphi}) and (\ref{Yphi}) are independent of the rotation angle $\phi$, and are exactly the same as those given in (\ref{XY-def}), $X(\phi)=X$ and $Y(\phi)=Y$. Therefore, $X$ and $Y$ are frame-invariant. Furthermore, the GFR angle is also frame-invariant.

\subsection{EMW case}
In this subsection, we present the main results of the GFR of EMWs for comparison. The unit transverse vector $\bm{\mathcal{E}}$ (\ref{E-def}) is seen as an electric vector of EMWs. The transformation of $\bm{\mathcal{E}}$ between source and detector positions is given by (\ref{E-trans}). $\bm{\mathcal{E}}$ is projected onto frame $(\hat{\mathbf{u}}, \hat{\mathbf{v}})$ as shown in Fig.\,\ref{fig:geometry figure} and is decomposed as $\bm{\mathcal{E}}=\mathcal{E}^{x}\hat{\mathbf{u}}+\mathcal{E}^{y}\hat{\mathbf{v}}$. These two polarization modes are given by
\begin{equation}
\label{E-projection}
\begin{aligned}
\left(\begin{array}{c}\mathcal{E}^{x}\\ \mathcal{E}^{y}\end{array}\right)
=\bm{\mathcal{N}}
\left(\begin{array}{c}
\mathcal{E}^{\theta} \\ 
\mathcal{E}^{\varphi}
\end{array}\right),\quad
\bm{\mathcal{N}}
\equiv\left(\begin{array}{cc}
\hat{u}^{\theta} & \hat{u}^{\varphi}\\
\hat{v}^{\theta} & \hat{v}^{\varphi}
\end{array}\right),
\end{aligned}
\end{equation}
where the $r$-component has been ignored. Similar to GWs, the polarization vector of EMWs has only two independent components $(\mathcal{E}^{\theta},\mathcal{E}^{\varphi})$. The matrix $\bm{\mathcal{N}}$ projects $(\mathcal{E}^{\theta},\mathcal{E}^{\varphi})$ onto two basis vectors, and gives two polarization modes $(\mathcal{E}^{x},\mathcal{E}^{y})$ of EMWs. Combining (\ref{E-projection}) and (\ref{E-trans}), the transformation between initial and final polarization modes is obtained
\begin{equation}
\label{ExEy-trans}
\begin{aligned}
\left(\begin{array}{c}\mathcal{E}^{x}_{d} \\ \mathcal{E}^{y}_{d} \end{array}\right)
=\bm{\mathcal{N}}_{d}
\left(\begin{array}{c}\mathcal{E}^{\theta}_{d}\\ \mathcal{E}^{\varphi}_{d}\end{array}\right)
=\bm{\mathcal{N}}_{d}\bm{\mathcal{T}}
\left(\begin{array}{c}\mathcal{E}^{\theta}_{s}\\ \mathcal{E}^{\varphi}_{s}\end{array}\right)
=\bm{\mathcal{N}}_{d}\bm{\mathcal{T}}\bm{\mathcal{N}}_{s}^{-1}
\left(\begin{array}{c}\mathcal{E}^{x}_{s}\\ \mathcal{E}^{y}_{s}\end{array}\right).
\end{aligned}
\end{equation}
The production of these three matrices reduces to a rotation matrix,
\begin{equation}
\label{EW-rotation}
\bm{\mathcal{N}}_{d}\bm{\mathcal{T}}\bm{\mathcal{N}}_{s}^{-1}
\equiv\left(\begin{array}{cc}\cos\chi_{e} & -\sin\chi_{e}\\ \sin\chi_{e} & \cos\chi_{e}\end{array}\right).
\end{equation}
This rotation matrix (\ref{EW-rotation}) relates the initial and final state of the lensed EMW. The change between the initial and final states is a rotation of the polarization vector in the $(\hat{\mathbf{u}}-\hat{\mathbf{v}})$ plane, that is, the GFR effect. In the rotation matrix (\ref{EW-rotation}), $\chi_{e}$ represents the GFR angle of the lensed EMWs. Compared with the GW case, the rotation angle of EMW is two times smaller than that of GWs since the photons are spin-1 particles. Eq.(\ref{EW-rotation}) gives the GFR angle $\chi_e$,
\begin{equation}
\label{chi-e}
\chi_{e}=\frac{1}{\sqrt{1+x^2}}\frac{X-x}{\sqrt{1+X^2}}.
\end{equation}
This result is different from the GFR angle of GWs (\ref{chi-g}). Interestingly, the GFR angle of the lensed EMWs and GWs have the same results under the WDL approximation.

\section{Results}
\label{sec-5}
This section aims to determine the value of $x$, $X$, and $Y$ and then predicts the GFR angle of GWs induced by a Kerr BH. The source, the lens BH, and the detector are well aligned in a strong lensing scenario. We define the angular shifts of the gravitons as
\begin{equation}
\label{shifts-def}
\Delta\theta\equiv\theta_{d}+\theta_{s}-\pi\quad\text{and}\quad \Delta\varphi\equiv\varphi_{d}-\varphi_{s}-\pi.
\end{equation}
These two quantities can be fully determined by the two motion constants, $\xi$ and $\eta$, and the final angular position of graviton relative to Kerr BH, $\theta_{d}$. In this study, as the impact parameter of graviton to the lens BH, $r_{m}$, is much larger than the gravitational radius of the BH, i.e. $\tilde{M}\equiv M/r_{m}^{(0)}, \tilde{a}\equiv a/r_{m}^{(0)}\sim\epsilon\ll1$, where $r_{m}^{(0)}=\sqrt{\eta+\xi^2}$ is the zero-order approximation of $r_{m}$. Thus, several quantities are small: $\Delta\theta, \Delta\varphi$, $r_{m}^{(0)}/r_{s}$, and $r_{m}^{(0)}/r_{d}$, which allows us to integrate geodesic equations in the WDL. The magnitudes of $r_{m}^{(0)}/r_{s}$ and $r_{m}^{(0)}/r_{d}$ are denoted by $\nu$, which is different from $\epsilon$. Without loss of generality, we assume the two have a similar order of magnitude, $\nu\sim\epsilon$.

We consider a static detector and a static GW source in the asymptotically flat region of a Kerr BH. The geometric relation between source and detector positions, $(r_{s},\theta_{s},\varphi_{s})$ and $(r_{d},\theta_{d},\varphi_{d})$, is derived from integrals
\begin{equation}
\label{r-theta-integral}
\int\frac{dr}{\pm\sqrt{R(r)}}
=\int\frac{d\theta}{\pm\sqrt{\Theta(\theta)}}
\end{equation}
and
\begin{equation}
\label{phi-integral}
\varphi_{d}-\varphi_{s}
=\int\frac{2Mar-a^2\xi}{\pm\Delta\sqrt{R(r)}}dr
+\int\frac{\xi\csc^2\theta}{\pm\sqrt{\Theta(\theta)}}d\theta,
\end{equation}
where the radial function $R(r)=r^4+(a^2-\xi^2-\eta)r^2+2Mr[\eta+(\xi-a)^2]-a^2\eta$, and the angular function $\Theta(\theta)=\eta+(a^2-\xi^2\csc^2\theta)\cos^2\theta$. The integral path is $r_{s}\rightarrow r_{m}\rightarrow r_{d}$ and $\theta_{s}\rightarrow \theta_{m}\rightarrow \theta_{d}$, with $r_{m}$ being the largest zero point of function $R(r)$, pericenter radius of gravitons paths. And $\theta_{m}=\theta_{\min/\max}$ is a root of $\Theta(\theta)=0$, which corresponds to the maximum or the minimum of the $\theta$ coordinate on the trajectory. Up to the order of $\epsilon^3$, the approximated solutions to $r_{m}$ and $\theta_{m}$ are
\begin{equation}
\label{r-m}
r_{m}\approx r^{(0)}_{m}
\left(1-\tilde{M}-\frac{3}{2}\tilde{M}^2+2\tilde{M}\tilde{a}\tilde{\xi}-\frac{1}{2}\tilde{a}^2\tilde{\xi}^2-4\tilde{M}^3+6\tilde{M}^2\tilde{a}\tilde{\xi}-2\tilde{M}\tilde{a}^2\tilde{\xi}^2\right)
\end{equation}
and
\begin{equation}
\label{theta-m}
\cos\theta_{m}\approx\pm\tilde{\eta}^{1/2}\left(1+\frac{1}{2}\tilde{a}^2\tilde{\xi}^2\right),
\end{equation}
where the dimensionless motion constants $\tilde{\xi}$ and $\tilde{\eta}$ are $\tilde{\xi}\equiv\xi/r_{m}^{(0)}$ and $\tilde{\eta}\equiv\eta/[r_{m}^{(0)}]^2$.

Following the method provided by \cite{Bray1986}, we perform the $r$-integral and $\theta$-integral (\ref{r-theta-integral}) up to $\epsilon^3$. The results are
\begin{equation}
\label{int-r}
\begin{aligned}
\int_{r_{s}}^{r_{d}}\frac{|dr|}{\sqrt{R(r)}}
&\approx\frac{1}{r_{m}^{(0)}}
\left(1+\frac{1}{2}\tilde{a}^2\right)
\left[\pi\left(1-\frac{3}{4}\tilde{a}^2\tilde{\eta}+\frac{15}{4}\tilde{M}^2-15\tilde{M}^2\tilde{a}\tilde{\xi}\right)\right.\\
&\left.+4\tilde{M}-8\tilde{M}\tilde{a}\tilde{\xi}+\frac{128}{3}\tilde{M}^3+10\tilde{M}\tilde{a}^2-16\tilde{M}\tilde{a}^2\tilde{\eta}\right]
-\frac{1}{r_{m}^{(0)}}\left(\mathcal{W}+\mathcal{Q}\right),
\end{aligned}
\end{equation}
and
\begin{equation}
\label{int-theta}
\int_{\theta_{s}}^{\theta_{d}}\frac{|d\theta|}{\sqrt{\Theta(\theta)}}
\approx\frac{1}{r_{m}^{(0)}}\left(1-\frac{3}{4}\tilde{a}^2\tilde{\eta}+\frac{1}{2}\tilde{a}^2\right)
\times\left\{\pi\mp\arctan\left[\left(1+\frac{1}{4}\tilde{a}^2\tilde{\eta}\right)\cot\sigma_{s}\right]\mp\arctan\left[\left(1+\frac{1}{4}\tilde{a}^2\tilde{\eta}\right)\cot\sigma_{d}\right]\right\},
\end{equation}
respectively. In (\ref{int-theta}), the rescaled angular coordinate is $\cos\sigma\equiv\cos\theta/|\cos\theta_{m}|$. The upper sign in (\ref{int-theta}) corresponds to the ray that passes through $\theta_{\min}$, and the lower sign corresponds to $\theta_{\max}$. The new symbols $\mathcal{W}$ and $\mathcal{Q}$ in (\ref{int-r}) are defined as
\begin{equation}
\label{W-Q-def}
\mathcal{W}=\frac{r_{m}^{(0)}}{r_{s}}+\frac{r_{m}^{(0)}}{r_{d}},
\quad\text{and}\quad
\mathcal{Q}=\frac{1}{6}\left[\left(\frac{r_{m}^{(0)}}{r_{s}}\right)^3+\left(\frac{r_{m}^{(0)}}{r_{d}}\right)^3\right].
\end{equation}
The last term in (\ref{int-r}) is absent in Ref. \cite{Ishihara1988} since they have assumed that the source and detector are far from the lens object, equivalently, $\epsilon\gg\nu$. 

Inserting the $r$-integral (\ref{int-r}) and $\theta$-integral (\ref{int-r}) into (\ref{r-theta-integral}), the relationship between $\theta_{d}$ and $\theta_{s}$ is obtaind,
\begin{equation}
\label{theta-S-D}
\cos\theta_{s}
\approx-\cos\delta\cos\theta_{d}
\mp\sin\delta\left\{\mu+\tilde{a}^2\left[\frac{1}{2\mu}\tilde{\eta}\tilde{\xi}^2+\frac{\mu}{4}\left(\tilde{\eta}-2\mu^2\right)\right]\right\},
\end{equation}
where $\mu$ is defined as $\mu=\sqrt{\tilde{\eta}-\cos\theta_{d}}$ and $\delta$ is
\begin{equation}
\label{delta}
\begin{aligned}
\delta=4\tilde{M}
+\frac{15\pi}{4}\tilde{M}^2
-8\tilde{M}\tilde{a}\tilde{\xi}
+\frac{128}{3}\tilde{M}^3
-15\pi\tilde{M}^2\tilde{a}\tilde{\xi}
-13\tilde{M}\tilde{a}^2\tilde{\eta}
+10\tilde{M}\tilde{a}^2
-\left(1-\frac{1}{2}\tilde{a}^2+\frac{3}{4}\tilde{a}^2\tilde{\eta}\right)\mathcal{W}-\mathcal{Q}.
\end{aligned}
\end{equation}
Except for the last two terms, this definition is completely consistent with that provided by \cite{Bray1986,Ishihara1988}. From (\ref{theta-S-D}) and (\ref{phi-integral}), the angular shifts of the gravitons (\ref{shifts-def}) now become
\begin{equation}
\label{Delta-theta}
\begin{aligned}
\Delta\theta\approx&\pm\mu\csc\theta_{d}\delta
-\frac{1}{2}\tilde{\xi}^2\cot\theta_{d}\csc^2\theta_{d}\delta^2\\
&\qquad\pm\frac{1}{4\mu}\csc\theta_{d}\left(2\tilde{\eta}\tilde{\xi}^2+\tilde{\eta}\mu^2-2\mu^4\right)\tilde{a}^2\delta
\pm\frac{\mu}{6}\delta^3\tilde{\xi}^2\csc^5\theta_{d}\left(2\mu^2+2\tilde{\xi}^2-3\right),
\end{aligned}
\end{equation}
and
\begin{equation}
\label{Delta-varphi}
\begin{aligned}
\Delta\varphi
&=4\tilde{M}\tilde{a}
-8\tilde{M}\tilde{a}^2\tilde{\xi} 
+5\pi\tilde{M}^2\tilde{a}
+\delta\tilde{\xi}\csc^2\theta_{d}
\pm\delta^2\mu\tilde{\xi}\cot\theta_{d}\csc ^3\theta_{d}
+\frac{1}{2}\tilde{a}^2\tilde{\xi}\delta\\
&\qquad+\frac{1}{4}\tilde{a}^2\delta
\Bigg[\left(-3\tilde{\eta}^3+4\tilde{\eta}^2\mu^2
+6\tilde{\eta}^2+\tilde{\eta}\mu^4
-6\tilde{\eta}\mu^2
-3\tilde{\eta}-2\mu^6+2\mu^2\right)\tilde{\xi}\csc^6\theta_{d}\\
&\qquad\qquad+\frac{1}{2}\left(-\tilde{\eta}^2+\tilde{\eta}\mu^2+\tilde{\eta}-2\mu^2\right)
\tilde{\xi}\csc^4\theta_{d}\Bigg]
+\frac{1}{3}\delta^3
\left(\tilde{\eta}^2+\tilde{\eta}\mu^2-\tilde{\eta}-2\mu^4+2\mu^2\right)
\tilde{\xi}\csc^6\theta_{d}.
\end{aligned}
\end{equation}
Approximately, we have $\Delta\theta$, $\Delta\varphi\sim\epsilon$ from (\ref{Delta-theta}) and (\ref{Delta-varphi}) \cite{Ishihara1988,Bray1986}. So far, we have completed the calculation of the null geodesics equations of gravitons. The angular shifts are determined by the dimensionless motion constants and detector position.

The quantities $x$ (\ref{x-def}), $X$, and $Y$ (\ref{XY-def}) are expressed in angular positions of the source and detector, $(\theta_{s},\varphi_{s})$ and $(\theta_{d},\varphi_{d})$, where $\varphi_{d}$ can be set as $0$ since Kerr BH has an axisymmetric spacetime. The quantities $x$, $X$, and $Y$ can then be expressed as functions of $\theta_{d}$ and angular shift, $\Delta\theta$ and $\Delta\varphi$. We use the angular shifts of gravitons, $\theta_{d}$, $\Delta\theta$ (\ref{Delta-theta}), and $\Delta\varphi$ (\ref{Delta-varphi}) for $x$ (\ref{x-def}), $X$, and $Y$ (\ref{XY-def}). Up to the order of $\epsilon^3$, they are given by
\begin{equation}
\label{xx}
\begin{aligned}
x&\approx\pm\Bigg[\frac{\tilde{\xi}}{\mu}\cot\theta_{d}\Delta\theta
+\frac{1}{\mu}\cos\theta_{d}\sin\theta_{d}\tilde{a}\Delta\theta
-\frac{1}{2}\tilde{\xi}\left(\frac{\tilde{\xi}^4}{\mu^3}-\frac{\tilde{\xi}^2}{\mu^3}+\frac{2\tilde{\xi}^2}{\mu}+\mu-\frac{2}{\mu}\right)\csc^2\theta_{d}\Delta\theta^2\\
&\qquad+\frac{1}{2}\left(\frac{\tilde{\xi}^3}{\mu^3}-\frac{\tilde{\xi}}{\mu^3}+\frac{\tilde{\xi}}{\mu}\right)\sin\theta_{d}\cos\theta_{d}\tilde{a}^2\Delta\theta
-\frac{1}{2}\left(\frac{\tilde{\xi}^4}{\mu^3}-\frac{\tilde{\xi}^2}{\mu^3}-\mu\right)\tilde{a}\Delta\theta^2\\
&\qquad\qquad-\left(\frac{\tilde{\xi}^5}{2\mu^5}-\frac{\tilde{\xi}^3}{2\mu^5}+\frac{\tilde{\xi}^3}{\mu^3}-\frac{\tilde{\xi}}{\mu^3}+\frac{\tilde{\xi}}{6\mu}\right)\cot\theta_{d}\Delta\theta^3\Bigg],
\end{aligned}
\end{equation}
\begin{equation}
\label{XX}
X\approx\cos\theta_{d}\Delta\varphi
+\frac{1}{2}\sin\theta_{d}\Delta\theta\Delta\varphi\\
+\frac{1}{12}\left(1+3\cos^2\theta_{d}\right)\cos\theta_{d}\Delta\varphi^3,
\end{equation}
and
\begin{equation}
\label{YY}
Y\approx\cos^2\theta_{d}\Delta\varphi^2+\sin\theta_{d}\cos\theta_{d}\Delta\theta\Delta\varphi^2.
\end{equation}
In $x$'s expression (\ref{xx}), the $\pm$ symbol comes from the definition of $\beta$. Gravitons pass the maximum or minimum position of $\theta$, corresponding to the upper or lower sign.

In (\ref{chi-g}) and (\ref{chi-e}), the GFR angle of lensed GWs and EMWs are expressed as functions of symbols $x$, $X$, and $Y$. Inserting (\ref{xx}), (\ref{XX}), and (\ref{YY}) into (\ref{chi-g}), up to $\epsilon^3$, the GFR angle is finally obtained,
\begin{equation}
\label{chi-g-result}
\chi_{g}=\tilde{a}\mathcal{W}\cos\theta_{d}
+\frac{5}{4}\pi\tilde{M}^2\tilde{a}\cos\theta_{d}
\pm\mu\left(2\tilde{M}\tilde{a}\mathcal{W}
-\frac{1}{2}\tilde{a}\mathcal{W}^2\right).
\end{equation}
The first term is of order $\epsilon^2$, while the last three are of order $\epsilon^3$. For $\nu\gtrsim\epsilon$, $\tilde{a}\mathcal{W}\cos\theta_{d}$ is the leading term of GFR angle. Since the dragging effect of Kerr BH causes GFR, the rotation angle is proportional to the angular momentum of the BH, where $\cos\theta_{d}$ accounts for the projection along the L.O.S. When $\cos\theta_{d}<0$ ($>0$), both the BH and the polarization plane rotate clockwise (anticlockwise). In the leading and subleading order of approximations, we find the same GFR angle of EMWs. For the $\nu\ll\epsilon$ case, the first third-order term, $(5/4)\pi\tilde{M}^2\tilde{a}\cos\theta_{d}$, will be the leading term, which has been given in Ref. \cite{Ishihara1988}. Additionally, our GFR angle (\ref{chi-g-result}) is frequency-independent because of geometric optical approximation. 

The magnitude of the GFR angle can be estimated from (\ref{chi-g-result}). We assume that the mass of a supermassive BH is $M\sim10^{9}M_{\odot}\sim10{\rm AU}$, angular momentum along L.O.S. is $a\cos\theta_{d}\sim0.9M$ and the BH-source distance is $r_{s}\sim10^{3}{\rm AU}$. Let the pericenter of the graviton path be as close as possible to the event horizon without exceeding the weak deflection limit, e.g., $\sim10^{2}{\rm AU}$. In this case, the condition $\nu\sim\epsilon$ is satisfied, and the GFR angle can reach $0.5\ {\rm deg}$, which is close to the detection sensitivity of the third-generation GW detector network to polarization angle $\psi$.

Additionally, substituting (\ref{xx}) and (\ref{XX}) into (\ref{chi-e}), we derive the GFR angle of EMWs after a similar calculation. The result of EMW's GFR angle is same as the GW case, i.e., $\chi_{e}=\chi_{g}$, which is the consequence of geometrical optics approximation. Ref. \cite{Ishihara1988} has studied such effect, but they only obtained the first third-order term of our result (\ref{chi-g-result}). Because they assume that , the parameter $\nu$ is much smaller than $\epsilon$, this assumption is not necessary for the real situation. By setting $\nu\ll\epsilon$ (equivalently, $\mathcal{W}\ll1$), we reproduce the same result in Ref. \cite{Ishihara1988}.

\section{Imprints in lensing GW waveform}
\label{sec-6}
There are some GFR imprints in gravitationally-lensed GW signals. In the geometric optical approximation, the lensing GW signal is calculated through
\begin{equation}
\label{lensingGW}
\tilde{h}^{+,\times}_{\rm len}(f)=F(f)\ \tilde{h}^{+,\times}_{\rm unlen}(f),
\end{equation}
where $\tilde{h}$ is the frequency-domain waveform, and $f$ is the frequency of GW without cosmological redshift. The lensed and unlensed waveform is denoted by subscripts `len' and `unlen'. The complex transmission factor $F(f)$ is a useful tool to describe the wave effects, amplitude enhancement, and phase shift of gravitationally-lensed GWs. The Kirchhoff integral is derived from the Huygens principle, and depends on the time delay of all possible paths of photons or gravitons. The transmission factor is  calculated from Kirchhoff integral \cite{Schneider1993,GuoXiao2020,LiangDai2018,Takahashi2003},
\begin{equation}
\label{Ff-def}
F(f)\equiv\frac{f(1+z_{L})}{i}\frac{d_{L}d_{S}}{d_{LS}}\theta_{E}^2\int d^2\bm{\xi}e^{i2\pi f(1+z_{L})\tau(\bm{\xi},\bm{\eta})},
\end{equation}
where $\bm{\xi}$ is the 2-dimensional coordinate on the lens plane and $\bm{\eta}$ is the 2-dimensional coordinate of the source object on the source plane. $d_{S}$, $d_{L}$, and $d_{LS}$ are the angular diameter distances from the observer to the source, to the lens, and from the lens to the source, respectively. In (\ref{Ff-def}), $\theta_{E}$ is the Einstein deflection angle, and $z_{L}$ is the cosmological redshift of the lens object. $\tau(\bm{\xi},\bm{\eta})$ is the time delay of a photon or graviton emitted from $\bm{\eta}$ passing through $\bm{\xi}$, and finally arriving at detector. The exact definition of time delay is given by \cite{Schneider1993,GuoXiao2020,LiangDai2018,Takahashi2003}
\begin{equation}
\tau(\bm{\xi},\bm{\eta})\equiv
\frac{d_Ld_S}{d_{LS}}\theta_{E}^2
\left[\frac{1}{2}(\bm{\xi}-\bm{\eta})^2-\phi(\bm{\xi})\right].
\end{equation}
$\phi(\bm{\xi})$ is called the lens potential and can be generally determined from the surface mass density of the lens object. The nonspinning point-like mass assumption is not necessary. The readers can find more details in \cite{Schneider1993}. For example, the Schwarzschild lens is considered in \cite{Takahashi2003} and the Kerr lens in \cite{Baraldo1999}. 

In the geometrical optics limit, the $j$-th ($j=1,2$ for point-mass lens) image position $\bm{\xi}_j$ is determined by the lens equation, $\bm{\nabla}_{\bm{\xi}}\tau(\bm{\xi},\bm{\eta})=0$. Therefore, the transmission factor $F(f)$ is contributed by each image and is expressed as
\begin{equation}
F(f)=\sum_{j}F_j(f)
=\sum_{j}\sqrt{|\mu_j|}e^{2\pi if\tau_{j}-i\pi n_{j}},
\quad\tau_j\equiv\tau(\bm{\xi}_{j},\bm{\eta})
\end{equation}
whereas the Morse indices are $n_j = 0, 1/2, 1$ when $\bm{\xi}_j$ is a minimum, saddle and maximum of the time-delay function $\tau(\bm{\xi},\bm{\eta})$, respectively. The magnifications $\mu_{j}$ is defined as $\mu_{j}\equiv[1/\det(\partial\bm{\xi}/\partial\bm{\eta})]_{\bm{\xi}=\bm{\xi}_j}$. Specially, the point-mass lens forms two images in the geometrical optics limit. For a point-mass lens, when $|\tau_{1}-\tau_{2}|$ is large enough, the amplitude and phase of $F_j(f)$, i.e., $\sqrt{|\mu_j|}$ and $2\pi if\tau_j-\pi n_j$, describe the amplitude enhancement and phase shift of lensed GW relative to that of the unlensed one. This framework describes amplitude enhancement, phase shift, interference, and diffraction. However, the polarization effect of the lensed GW is not included. If taking the GFR effect into consideration, transformation (\ref{lensingGW}) becomes
\begin{equation}
\label{lensingGW-GFR}
\begin{aligned}
\tilde{h}^{+}_{j,{\rm len}}(f)
&=F_j(f)\left[\tilde{h}^{+}_{j,{\rm unlen}}(f)\cos2\chi_{j}
-\tilde{h}^{\times}_{j,{\rm unlen}}(f)\sin2\chi_{j}\right],\\
\tilde{h}^{\times}_{j,{\rm len}}(f)
&=F_j(f)\left[\tilde{h}^{+}_{j,{\rm unlen}}(f)\sin2\chi_{j}
+\tilde{h}^{\times}_{j,{\rm unlen}}(f)\cos2\chi_{j}\right],\\
\end{aligned}
\end{equation}
or equivalently
\begin{equation}
\label{lensingGW-GFR-LR}
\tilde{h}^{L,R}_{j,{\rm len}}(f)=F_{j}(f)\ \tilde{h}^{L,R}_{j,{\rm unlen}}(f)
\exp\left(-i\cdot2\chi_{j}\right),
\end{equation}
in terms of left-hand and right-hand polarization modes. We abbreviate the GFR angle of $j$-th image as $\chi_{j}$. The left-hand and right-hand polarization modes are defined as $h^{L,R}\equiv (1/\sqrt{2})(h^{+}\pm ih^{\times})$ \cite{Maggiore2008}. As shown in (\ref{lensingGW-GFR}) or (\ref{lensingGW-GFR-LR}), the lensing effects can be divided into two parts. Firstly, the unlensed frequency-domain GW waveform will be multiplied by a complex factor $F(f)$, which contains amplitude enhancement and phase shift effects. Secondly, the polarization plane rotates by an angle $\chi_{g}$ caused by the GFR effect. Our work gives the theoretical prediction of the GFR angle (\ref{chi-g-result}). 

The unlensed signal received by detectors can be written as a linear combination of $+$ and $\times$ polarization modes,
\begin{equation}
\label{unlensed-signal}
\tilde{h}_{\rm unlen}
=F_{+}(\alpha,\lambda,\psi)\tilde{h}_{\rm unlen}^{+}
+F_{\times}(\alpha,\lambda,\psi)\tilde{h}^{\times}_{\rm unlen}.
\end{equation}
$F_{+,\times}(\alpha,\lambda,\psi)$ is the pattern function of GW detectors \cite{Maggiore2008} for $+$ and $\times$ modes. $(\alpha,\lambda)$ are the right ascension and declination of GW sources, and $\psi$ is the so-called polarization angle. The polarization angle is defined in terms of the direction of orbital angular moment of binary $\hat{\mathbf{L}}$ and GW propagation $\hat{\mathbf{N}}$ \cite{Jaranowski1998}. However, this definition is inconvenient for lensing system. Instead, we define the polarization angle using the basis vectors $(\hat{\mathbf{u}},\hat{\mathbf{v}})$ shown in Fig. \ref{fig:geometry figure}. Thus, the lensed GW signal is given by
\begin{equation}
\label{lensing-signal}
\begin{aligned}
\tilde{h}_{j,{\rm len}}
&=\tilde{h}_{j,{\rm len}}^{+}
F_{+}(\alpha'_j,\lambda'_j,\psi)
+\tilde{h}_{j,{\rm len}}^{\times}
F_{\times}(\alpha'_j,\lambda'_j,\psi)\\
&=F_j(f)\Big\{
(\tilde{h}_{j,{\rm unlen}}^{+}\cos2\chi_{j}
-\tilde{h}_{j,{\rm unlen}}^{\times}\sin2\chi_{j})
[F_{+}(\alpha'_j,\lambda'_j,0)\cos2\psi
-F_{\times}(\alpha'_j,\lambda'_j,0)\sin2\psi]\\
&\qquad+(\tilde{h}_{j,{\rm unlen}}^{+}\sin2\chi_{j}
+\tilde{h}_{j,{\rm unlen}}^{\times}\cos2\chi_{j})
[F_{+}(\alpha'_j,\lambda'_j,0)\sin2\psi
+F_{\times}(\alpha'_j,\lambda'_j,0)\cos2\psi]\Big\}\\
&=F_j(f)\Big\{
\tilde{h}_{j,{\rm unlen}}^{+}
[F_{+}(\alpha'_j,\lambda'_j,0)\cos2(\psi-\chi_{j})
+F_{\times}(\alpha'_j,\lambda'_j,0)\sin2(\psi-\chi_j)]\\
&\qquad
+\tilde{h}_{j,{\rm unlen}}^{\times}
[F_{+}(\alpha'_j,\lambda'_j,0)\sin2(\psi-\chi_j)
F_{\times}(\alpha'_j,\lambda'_j,0)\cos2(\psi-\chi_{j})]
\Big\}\\
&=F_j(f)\left[
\tilde{h}_{j,{\rm unlen}}^{+}
F^{+}(\alpha'_j,\lambda'_j,\psi-\chi_{j})
+\tilde{h}_{j,{\rm unlen}}^{\times}
F^{+}(\alpha'_j,\lambda'_j,\psi-\chi_j)\right].
\end{aligned}
\end{equation}
In the second step, we use the properties of pattern functions, $F_{+}(\alpha,\lambda,\psi)=F_{+}(\alpha,\lambda,0)\cos2\psi-F_{\times}(\alpha,\lambda,0)\sin2\psi$ and $F_{\times}(\alpha,\lambda,\psi)=F_{+}(\alpha,\lambda,0)\sin2\psi+F_{\times}(\alpha,\lambda,0)\cos2\psi$. The massive lens object deflects the path of gravitons, changing the direction of the incoming GW signal from $(\alpha,\lambda)$ to $(\alpha'_j,\lambda'_j)$ on the sky plane. From (\ref{lensing-signal}) we can find that the effect of GFR is equivalent to a change in the polarization angle $\psi$ in the pattern function, $\psi\rightarrow\psi-\chi_{g}$. Additionally, the total amplitude of the lensed GW signal at detectors is proportional to $|F_j|/D_{L}$, where $D_{L}$ is the luminosity distance of the GW source. The total phase is $\phi_{c}+\arg(F_j)$, where $\phi_{c}$ is the coalescence phase of the binary system. Therefore, lensing amplification and the GW source distance are degenerate, the phase shift and initial phase $\phi_{c}$ are degenerate \cite{Ezquiaga2021}. Similarly, the GFR angle and initial polarization angle $\psi$ are degenerate. 

We need to find a way to break the degeneracy between the initial polarization angle and the GFR angle. In the geometrical optics limit, the lens focuses the original GW from several paths toward an observer, forming multiple images \cite{Baraldo1999,Sereno2006}. Every path's pericenter radius $r_{m}$ is usually different. The observed GFR effects from each image are distinguished.  A Kerr lens forms two images on the lens plane in the geometrical optical approximation. The observed values of the polarization angle of the lensed GW from each image will be $\psi_{j}=\psi-\chi_{j}$. These two GFR angles are different because of different graviton paths. Therefore, the difference in GFR angles is obtained,
\begin{equation}
\Delta\chi_{g}=\chi_{2}-\chi_{1}=\psi_{2}-\psi_{1}.
\end{equation}
From the GFR angle (\ref{chi-g-result}), we find that $\Delta\chi_{g}$ is a third-order quantity, $\sim\mathcal{O}(\epsilon^{3})$. The first term in (\ref{chi-g-result}), $\tilde{a}\mathcal{W}\cos\theta_{d}=a(1/r_{s}+1/r_{d})\cos\theta_{d}$, is independent of the pericenter radius $r_{m}^{(0)}$. The second-order corrections cancel each other out in $\Delta\chi_{g}$, and all of the three three-order terms survive. Therefore, by observing the images of lensed GW rays, one can obtain $\Delta\chi_{g}$ and estimate the spin of the lens black hole accordingly. However, such a method brings more difficulties. Firstly, it requires sophisticated modeling of gravitational lensing. Secondly, it makes the leading-order term vanish, and one has to compare the tiny difference between multiple images, which requires highly accurate measurement. We consider a supermassive BH and a GW source system similar to that in the previous section. The GW signals form two images on the lens plane. The impact distances of these two graviton orbits are assumed to be $10^2{\rm AU}$ and $1.1\times10^2{\rm AU}$. Generally, the difference between the two impact distances should not be too large because the GW rays with a significant impact distance will be too dim to be detected. Only considering the first three-order term, the subleading terms of the GFR angle of the two images are about $0.2\ {\rm deg}$ and $0.15\ {\rm deg}$, respectively. As a result, the GFR angle difference is $0.05\ {\rm deg}$, which is beyond the detectability of the third-generation GW detector or networks.

There are other potential methods to address this problem, such as considering the motion of lens objects or background GW sources. One can consider that the initial polarization angle of GWs does not change with the motion of the source. However, the propagation path of the GW and, thus, the GFR angle changes, when the lens object or GW source moves. Therefore, it is possible to detect the polarization angle changing with time from the lensed GW signals to obtain characteristics of the GFR effect. However, this method needs to consider the moving GW source and lens object, and solving the geodesic equation will become more challenging.

The actual measurement of the GFR angle helps reveal the relevant characteristics of spinning objects. Breaking the degeneracy is of great significance for the observation of GFR. More works are needed in this field for in-depth discussion.

\section{Conclusion and Discussion}
\label{sec-7}
The gravitationally-lensed GWs can be successfully described by a transmission factor, evaluated from the Kirchhoff integral \cite{Takahashi2003,Meena2019,Born1959,Baraldo1999,GuoXiao2020}. This framework includes the effects of amplitude enhancement, phase shift, interference, and diffraction, but the GFR effect is absent. In this work, by applying the Walker-Penrose theorem, we derive an explicit result of the GFR angle of GWs lensed by a Kerr BH (\ref{chi-g-result}) in the geometric optical approximation and WDL assumption. 

The GFR effect describes the mixture of original GW polarization modes, is an important correction to the polarization angle of gravitationally-lensed GW signals. It is a frame-dragging effect sourced by the angular momentum of spinning lens objects. The GFR of the lensed GW was not considered in previous works because of its weak effect. However, in this work, we find that the GFR angle is a second-order correction of the polarization angle rather than a third-order correction, as predicted in previous work. The GFR angle of the GW signal lensed by a central supermassive BH in a galaxy has been estimated in Section \ref{sec-5} ($\sim0.5\ {\rm deg}$). On the other hand, Fisher estimation for the standard deviation of polarization angle $\Delta\psi$ of the stellar-mass binary is about $0.6\ {\rm deg}$ for HLV/ET/B-DECIGO network \cite{Grimm2020}. This kind of system has been predicted by \cite{Graham2020}, and its GFR angle $\chi_{g}$ is close to the detection capability of the third-generation GW detector network. However, we have to break the degeneracy between the GFR and polarization angles.

As well known, the amplitude enhancement factor of lensed GWs and the luminosity distance of the GW source are degenerate, the phase shift and the coalescence phase of compact binary coalescence of lensed GWs are degenerate \cite{Ezquiaga2021}. Similarly, the GFR angle and initial polarization angle $\psi$ are also degenerate. Compared with an unlensed signal, only the polarization angle changes from $\psi$ to $\psi-\chi_{g}$ for lensed signals. Theoretically, one can break the degeneracy between the GFR and the initial polarization angle from the multiple lensed images, although higher-accuracy observations  is required.

The GFR effect is sourced from the angular momentum of spinning lens objects and changes the initial polarization state of lensed GW signals. As a result, this study may help future research measure the spin of BHs. Faraday rotation of EMW depends on both the magnetic field and the rotating gravitational field during propagation. However, the GFR of GW is only affected by the gravitational field. Therefore, this effect is suitable for measuring the BH or galaxy spin. Similarly, lensed EMWs and GWs from a binary system with an electromagnetic counterpart provide a potential method to measure the magnetic field along the line of sight. It is anticipated that many lensed GW signals will be detected in the future \cite{ShunSheng2018,Piorkowska2013,Hannuksela2019,Lo2021}, which will help measure the spin of more supermassive BHs and galaxies statistically and test the Universe's chiral symmetry \cite{HarranYu2020}.

~

\emph{Acknowledgments.}
We would like to thank Youjun Lu, Yefei Yuan, Xilong Fan, Shaoqi Hou and Aoxiang Jiang for helpful discussions and comments. This work is supported by the National Key R\&D Program of China Grant No. 2021YFC2203100, NSFC No. 12273035, 11773028, 11633001, 11653002, 11903030, 11873006 the Fundamental Research Funds for the Central Universities under Grant No. WK2030000036 and WK3440000004, Key Research Program of the Chinese Academy of Sciences, Grant No. XDPB15, and the science research grants from the China Manned Space Project with No.CMS-CSST-2021-B01, CMS-CSST-2021-B11 and CMS-CSST-2021-A12.

\bibliographystyle{apsrev4-2}
\bibliography{ref}

\end{document}